\documentclass[article,fleqn,final,authoryear,3p,times,preprint,preview]{elsarticle}
\usepackage{amsmath, amssymb, amsfonts}
\usepackage[export]{adjustbox}
\usepackage{subfig}
\usepackage{color, array, bm}
\usepackage{hyperref,graphicx}
\usepackage[]{units}
\usepackage{enumitem}
\usepackage{booktabs}
\usepackage{caption}
\usepackage[dvipsnames]{xcolor}
\hypersetup{colorlinks=true, citecolor=blue, linkcolor= magenta}
\usepackage[T1]{fontenc}
\usepackage{ae,aecompl}
\usepackage{txfonts,pifont}
 
\usepackage{mathptmx}

\newcommand{\src }{XTE~J1810$-$197 }
\newcommand\Tstrut{\rule{0pt}{2.6ex}}         
\definecolor{darkred}{rgb}{0.55, 0.0, 0.0}

\journal{New Astronomy}

\begin{document}
\begin{frontmatter}
\begin{center}
{\LARGE \bf Variable Absorption Line in \src
}
\\ 
\vspace{0.5cm}
Eda Vurgun$^{\mathrm{1^{\color{blue} \bigstar}}}$,
Manoneeta Chakraborty$^{\mathrm{2,3}}$,
Tolga G\"uver$^{\mathrm{4,5}}$ and \\
Ersin G{\"o}{\u g}{\"u}{\c s}$^{\mathrm{2}}$
\\
\vspace{0.5cm}
{ \it
$^{1}$\.{I}stanbul University, Graduate School of Science and Engineering, Department of Astronomy and Space Sciences, 34116, Beyaz\i t, \.{I}stanbul, Turkey\\
$^{2}$Sabanc\i~University, Faculty of Engineering and Natural
Sciences, Tuzla, 34956, \.{I}stanbul, Turkey \\
$^{3}$Center of Astronomy, Indian Institute of Technology Indore, Khandwa Road, Simrol Indore 453552, India \\
$^{4}$\.{I}stanbul  University,  Science Faculty,  Department  of
  Astronomy and Space Sciences, 34119, Beyaz\i t, \.{I}stanbul, Turkey 
  \\
$^{5}$\.{I}stanbul University, Observatory Research and Application Center, Istanbul University, 34119, \.{I}stanbul, Turkey}
\begin{footnotesize}
\color{blue}
\vspace{0.2cm}
$^{\bigstar}{E-mail: eda.vurgun@ogr.iu.edu.tr}$
\end{footnotesize}
\small\itshape
\end{center}
\vspace{-2.5cm}
\begin{abstract}

We report the results of a long-term spectral and timing study of the 
first transient magnetar, \src which was discovered  in 2003, when its X-ray luminosity increased $\approx$100 fold. We fit X-ray spectra of all archival X-ray observations using a two-component blackbody model, where the cool component is most likely originating from the whole surface of the neutron star and the hot component is from a much 
smaller hot spot. We investigate the long-term evolution of the surface 
emission characteristics via tracing its surface temperature, apparent 
emitting area and pulsed fraction. We evaluate the pulsed fraction in 
two energy intervals ($<$ 1.5 keV and $>$1.5 keV)  and show  that the 
\src exhibits slightly higher pulsed emission at energies above 1.5 keV. 
We explore the characteristics of an absorption line detected around  1.1~keV. We find that the absorption feature is highly variable and its profile is asymmetric. To accurately represent this feature, we introduced an asymmetric Gaussian profile, and quantified the level of asymmetry of the absorption feature. 

\end{abstract}

\end{frontmatter}


\section{Introduction}
Anomalous X-ray Pulsars  (AXPs) and Soft-Gamma Repeaters  (SGRs) are a
group  of neutron  stars, which  differ from  others because  of their
numerous observational properties. These  objects have relatively long
spin  periods   ($>$2~s)   and  high  inferred   period  derivatives
($10^{-13}-10^{-11}$~s/s)  leading to  dipole magnetic  fields at  the
surface, of  the order  of $10^{14-15}$~G. These  temporal properties,
along with the observed energetic  short X-ray bursts and giant flares
from  these objects  led  to the  suggestion that  AXPs  and SGRs  are
magnetars  \citep{Duncan1992, 1996};  objects  whose persistent  X-ray
luminosity arises from the decay of a superstrong magnetic field.

Although  the  first discovered  AXPs  and  SGRs were  persistent  and
relatively  bright X-ray  sources,  today the  McGill Online  Magnetar
Catalog{\footnote{http://www.physics.mcgill.ca/~pulsar/magnetar/main.html}}
\citep{Olausen2014}  consists  of nine  transient  objects  out of  23
confirmed AXPs and  SGRs. All known magnetars exhibit a  wide range of
temporal    variations,     such    as    bursts,     glitches,    and
outbursts. Transient Anomalous X-ray  Pulsars (TAXPs) display distinct
bursts and/or outbursts, which are characterized by an abrupt increase
in X-ray flux, followed by a long-term flux decay \citep{Kaspi2007}.

\src is the  prototypical TAXP, which was first discovered  in 2003 by
\citet{Ibrahim2004}, when  its X-ray luminosity increased  suddenly by
$\approx$100    times     compared    to    its     quiescent    state
\citep{Halpern2005}. Since its discovery, the X-ray flux of the source
declined exponentially  and it  is now  thought to  be in  a quiescent
phase \citep{Gotthelf2007}.  \src has a  spin period of  5.54~s period
and      a      large       period      derivative      $10^{-11}$~s/s
\citep{Ibrahim2004}. These  timing properties  imply a  surface dipole
magnetic   field   strength    of   $2-3\times10^{14}$~G,   which   is
independently confirmed  by \citet{Guver2007} using an  X-ray spectral
model that take  into account the radiative processes  taking place in
the atmosphere of a  magnetar and its magnetosphere. \citet{Guver2007}
also found that the X-ray spectra of the source obtained with XMM-{\it
  Newton}  until 2006  could be  well modeled  with the  cooling of  a
roughly constant area hot spot. 

Thanks to  extensive observing campaigns performed  with Chandra X-ray 
Observatory and XMM-{\it Newton} the  spectral and timing evolution of
\src   has   been   followed   in    a   unique   and   detailed   way
\citep{Alford2016,Pintore2016}.  \citet{Alford2016} modeled  the X-ray
spectra  with two  or three  blackbody components,  where the  coldest
component is  held fixed throughout  all the observations.  They found
that the hot  spot remains on the surface with  a 0.3~keV temperature,
and   hence  the   pulsations.   Similar  to   \citet{Bernardini2009},
\citet{Alford2016}  report  the  presence  of a  spectral  feature  at
1.2~keV at all epochs, which if  assumed to be a proton cyclotron line
indicates a magnetic field strength of approximately $2\times 10^{14}$
G,   in  agreement   with  earlier   conclusions.  \citet{Pintore2016}
investigated  the  evolution  of  the pulsar  spin  period  and  found
evidence for two distinct regimes.  During the decay from the outburst
$\dot{\nu}$  was found  to be  highly variable;  only about  3000 days
after its outburst  onset it was possible to phase  connect. They also
report a possible  anti-glitch at around MJD 55400.  In another recent
study  on \src,  \citet{Camilo2016} reported  timing and  polarimetric
observations of the source using the Green Bank, Nan\c{c}ay and Parkes
radio telescopes. Radio pulsation was detected during the observations
performed between  2006$-$2008, which coincide  with the times  of the
frequency  change  reported  by   \citet{Pintore2016}.  In  this  time
interval,  unlike ordinary  radio pulsars  the pulsation  had a  large
day-to-day  fluctuation, which  resulted  in a  steep radio  spectrum,
rather  than a  flat-spectrum. \citet{Camilo2016}  also reported  that
\src has not been detected in the radio band since 2008. 

In this  paper we report on  the spectral and timing  evolution of the
\src spanning more than 12 years. Our systematic investigation focuses
on modeling its spectral continuum with two blackbody functions, and a
better  understanding  of the  spectral  feature  at 1.2~keV  with  an
asymmetric Gaussian fuction. We describe the observations of \src used
in our  analysis in  \S\ref{sec2}.  In \S\ref{sec3} and  \S\ref{sec4} 
we  present the details  of our  temporal  and  spectral analysis. 
In \S\ref{sec5}, we present the results of absorption line structure. 
In  \S\ref{sec6},  we discuss the implications of our results.


\section{Observations and Data Analysis}
\label{sec2}
We employed  34 archival observations  of \src collected  with Chandra
X-ray Observatory and XMM-{\it Newton}. \autoref{observations} lists
the log  of these  observations. We  present the  details of  our data
reduction methodology for each instrument separately.

\begin{table*}[t]
\centering
\caption{Log of archival X-ray observations.}
\label{observations}
\setlength{\tabcolsep}{5.4pt}
\scriptsize{
\begin{tabular}{ccccccc}
\hline \Tstrut
  \# & Observatory & Observation ID & Start Time & Date & Exp. Time & Data Mode\\
  & & & MJD & UT & ks & \\
\hline\\
1&	XMM-{\it Newton} &0161360301 & 52890.55714 & 2003 Sep 8  & 6.1  &	Small Window\\
2&	XMM-{\it Newton} &0161360401 & 52890.70589 & 2003 Sep 8  & 0.3  &	Small Window\\
3&	XMM-{\it Newton} &0161360501 & 53075.49562 & 2004 Mar 11 & 8.9  &	Large Window\\
4&	XMM-{\it Newton} &0164560601 & 53266.49812 & 2004 Sep 18 & 20.9 &	Full Window\\
5&	XMM-{\it Newton} &0301270501 & 53447.99696 & 2005 Mar 18 & 30.5 &	Large Window\\
6&	XMM-{\it Newton} &0301270401 & 53633.44414 & 2005 Sep 20 & 26.5 &	Large Window\\
7&	XMM-{\it Newton} &0301270301 & 53806.79024 & 2006 Mar 12 & 19.2 &	Large Window\\
8&	Chandra 		& 6660	     & 53988.80975 & 2006 Sep 10 & 27.4 &	VFAINT\\
9&	XMM-{\it Newton} &0406800601 & 54002.06184 & 2006 Sep 24 & 37.9 &	Large Window\\
10&	XMM-{\it Newton} &0406800701 & 54165.77228 & 2007 Mar 6  & 32.9 &	Large Window\\
11&	XMM-{\it Newton} &0504650201 & 54359.06106 & 2007 Sep 16 & 67.0 &	Large Window\\
12&	Chandra 		& 7594	     & 54543.00393 & 2008 Mar 18 & 26.9 &	VFAINT\\
13&	XMM-{\it Newton} &0552800201 & 54895.65530 & 2009 Mar 5  & 39.8 &	Large Window\\
14&	XMM-{\it Newton} &0605990201 & 55079.62303 & 2009 Sep 5  & 17.9 &	Large Window\\
15&	XMM-{\it Newton} &0605990301 & 55081.55238 & 2009 Sep 7  & 16.3 &	Large Window\\
16&	XMM-{\it Newton} &0605990401 & 55097.70535 & 2009 Sep 23 & 11.0 &	Large Window\\
17&	Chandra			 &11102	     & 55136.66053 & 2009 Nov 1  & 22.8 &	VFAINT\\
18&	Chandra 		& 12105	 	 & 55242.69038 & 2010 Feb 15 & 11.4 &	VFAINT\\
19&	Chandra 		& 11103	 	 & 55244.74581 & 2010 Feb 17 & 11.5 &	VFAINT\\
20&	XMM-{\it Newton} &0605990501 & 55295.18394 & 2010 Apr 9  & 7.1  &	Large Window\\
21&	Chandra 		& 12221	 	 & 55354.13126 & 2010 Jun 7  & 9.5  &	FAINT\\
22&	XMM-{\it Newton} &0605990601 & 55444.67699 & 2010 Sep 5  & 8.4  &	Large Window\\
23&	Chandra 		& 13149	 	 & 55494.16719 & 2010 Oct 25 & 14.6 &	FAINT\\
24&	Chandra			 &13217	  	 & 55600.99240 & 2011 Feb 8  & 13.6 &	FAINT\\
25&	XMM-{\it Newton} &0671060101 & 55654.08608 & 2011 Apr 3  & 14.3 &	Large Window\\
26&	XMM-{\it Newton} &0671060201 & 55813.38491 & 2011 Sep 9  & 11.8 &	Large Window\\
27&	Chandra 		& 13746	 	 & 55976.37662 & 2012 Feb 19 & 18.2 &	VFAINT\\
28&	Chandra 		& 13747	 	 & 56071.35995 & 2012 May 24 & 18.2 &	VFAINT\\
29&	XMM-{\it Newton} &0691070301 & 56176.98015 & 2012 Sep 6  & 14.6 &	Large Window\\
30&	XMM-{\it Newton} &0691070401 & 56354.19799 & 2013 Mar 3  & 7.4  &	Large Window\\
31&	XMM-{\it Newton} &0720780201 & 56540.85575 & 2013 Sep 5  & 18.8 &	Large Window\\
32&	Chandra 		& 15870	 	 & 56717.31190 & 2014 Mar 1  & 18.2 &	VFAINT\\
33&	XMM-{\it Newton} &0720780301 & 56720.97152 & 2014 Mar 4  & 19.6 &	Large Window\\
34&	Chandra 		& 15871	     & 56907.94909 & 2014 Sep 7  & 18.2 &	VFAINT\\
  \hline 
\end{tabular}}
\end{table*}

\subsection{Chandra X-ray Observatory}

There are  a total of  14 Chandra observations performed  between 2006
and  2015. Out  of  these, 12  were performed  with  the Advanced  CCD
Imaging Spectrometer  (ACIS) and  2 of them  were performed  using the
High  Resolution Camera  (HRC). Since  the HRC  does not  provide fine
spectral  resolution,  we  employed   only  the  Chandra  observations
performed with ACIS.  We used CIAO and CALDB version  4.5, 4.5.5.1 for
the calibration of the data. For  the last two observations (15870 and
15871), we used  CIAO and CALDB version 4.6, 4.6.3.  We extracted each
spectrum  using  the {\it  specextract}  tool  following the  standard
Chandra data  analysis threads with  a 6 arcsec extraction  radius for
source and  7 arcsec  for background  regions. Response  and ancillary
response  files were  generated  using the  {\it  mkacisrmf} and  {\it
  mkarf} tools, respectively. We grouped the resulting spectra so that
each spectral channel contains a minimum  of 50 counts. Often the time
resolution  provided  by the  Chandra  data  was $\sim  0.4$~s,  which
usually correspond to 1/8 sub-array mode of ACIS.

\subsection{XMM-{\it Newton}}

We inspected all 22 X-ray  datasets obtained with the EPIC-pn detector
between  2003  and  2014.  We  did  not  employ  one  exposure  (ID  :
0552800301) because  of its high  solar
particle background.  We performed the  calibration of the  data using
the Science Analysis Software (SAS) version 12.0.1. The calibrated and
cleaned event files were created using  the {\it epproc} task. We used
the {\it  rmfgen} and {\it  arfgen} tools  within SAS to  generate the
response files.  All X-ray spectra  were grouped  to have at  least 50
counts  in  each  spectral  energy  bin  and  not  to  oversample  the
instrumental energy resolution  by more than a factor  of three, using
the {\it specgroup} tool. The  time resolution offered by EPIC-pn data
is 48  ms and  73 ms for  the large window  and full  frame sub-modes,
respectively.


\section{Temporal Analysis}
\label{sec3}
\begin{figure*}[t]
  \begin{centering}
 \includegraphics[scale=0.3]{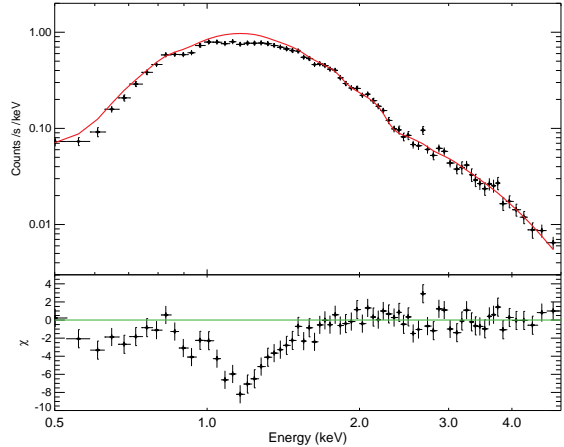}
 \vspace{-1.5cm}
 \caption{{\it Upper  panel:} XMM-Newton  spectrum of \src  as observed
 starting  from  MJD  53633.44414,  and  the best  fit  model  of  two
 blackbody functions (solid line). {\it Lower panel:} The residuals of the fit.}
 \label{fig:xmm}
 \end{centering}
\end{figure*}

For  the timing  analysis, we  investigated the  timing features  like
pulsed    fraction   for    the   observations    listed   in \autoref{tab:pulse}.  
We  extracted  the  photon  arrival  times  for  the XMM-{\it   Newton}  and   
Chandra  data   and  performed   barycentric corrections  on   them  using  the  
position   coordinates  from  VLBA observations \citep{Helfand2007}.  A long term 
phase  connected timing solution could not be obtained for this source due to data 
constraints \citep{Pintore2016, Camilo2016}.  Therefore, for each  observation, we
calculated  the  local spin  frequency  of  pulsar using  the  $Z_n^2$
statistic \citep{Buccheri1983}. This statistic is defined as
\\
\begin{eqnarray}
\label{eq:1}
Z_n^2=(2/N)\Sigma^n_k( (\Sigma^N_j \cos k\phi_j)^2 + (\Sigma^N_j \sin k\phi_j)^2)
\end{eqnarray}\nonumber
\\
where $\phi$ is the phase array ,  N is the number of trial phase bins
and  n  is  the  number  of harmonics.  We  created  a  $Z_n^2$  power
distribution for a  number of trial frequencies  ranging from 1.0/5.55
to 1.0/5.53  Hz with frequency  steps of  $10^{-7}$ Hz and  taking the
value of  $n$ to be 2.  The phases were calculated  by multiplying the
barycenter corrected times by each trial frequency, which were used to
compute  the  $Z_n^2$ value  for  each  frequency. We  identified  the
frequency corresponding to the peak  of the $Z_n^2$ power distribution
as  the   local  spin  frequency  corresponding   to  that  particular
observation. The obtained frequencies match closely with that reported
in the  literature \citep{Camilo2016}. Using this  estimated frequency
for  each  observation, we  folded  the  data  at this  frequency  and
obtained  the  pulse  profiles.  In  order  to  examine  the  temporal
evolution with energy, the profiles were obtained in two energy ranges
0.5-1.5 and  1.5-10.0 keV (as chosen  based on the BB  components from
the spectral analysis). From the profiles in the two energy ranges, we
subtracted  the  background  contribution  and  calculated  the  pulse
fraction. We  defined the pulsed fraction as 
\\
\begin{eqnarray}
\label{eq:2}
PF=((1/N)*(\Sigma^N_i(R_i-R_{avg})^2 - \Delta R_i^2))^{1/2}/R_{avg}
\end{eqnarray}
\\
where  $R_i$ is  the count  rate in  i$^{th}$ phase  bin in  the pulse
profile, $\Delta  R_i$ is  the corresponding  error, $R_{avg}$  is the
average count rate in  the pulse profile and N is  the number of phase
bins.  Following the  above procedure, we obtained  the time evolution
of the  rms pulsed fraction in  two energy ranges using  both XMM-{\it
  Newton} and Chandra data which is  shown in the last panel of
\autoref{fig:xmm} and \autoref{tab:pulse}.

\begin{table*}[h]
\centering
 \caption{Pulsed fractions of \src in two energy bands}
 \label{tab:pulse}
\setlength{\tabcolsep}{5.4pt}
\scriptsize{
 \begin{tabular}{ccc}
  \hline \Tstrut
  \# & Pulsed Fraction & Pulsed Fraction\\
 & (0.5-1.5 keV) &    (1.5-10.0 keV) \\
\hline  \\
1&	  0.319 ${\color{darkred}\pm}$  0.007 &0.408 ${\color{darkred}\pm}$   0.005\\
2&	  0.349 ${\color{darkred}\pm}$  0.018 &0.537 ${\color{darkred}\pm}$   0.015\\
3&	  0.239 ${\color{darkred}\pm}$  0.008 &0.329 ${\color{darkred}\pm}$   0.006\\
4&	  0.228 ${\color{darkred}\pm}$  0.006 &0.342 ${\color{darkred}\pm}$   0.005\\
5&	  0.199 ${\color{darkred}\pm}$  0.006 &0.298 ${\color{darkred}\pm}$   0.006\\
6&	  0.149 ${\color{darkred}\pm}$  0.009 &0.222 ${\color{darkred}\pm}$    0.011 \\
7&	  0.154 ${\color{darkred}\pm}$  0.012 &0.235 ${\color{darkred}\pm}$    0.018\\
8&	  0.127 ${\color{darkred}\pm}$  0.016 &0.257 ${\color{darkred}\pm}$    0.025     \\
9&	  0.157 ${\color{darkred}\pm}$  0.009 &0.276 ${\color{darkred}\pm}$    0.016\\
10&   0.123 ${\color{darkred}\pm}$  0.011 &0.255 ${\color{darkred}\pm}$    0.028\\
11&   0.157 ${\color{darkred}\pm}$  0.007 &0.251 ${\color{darkred}\pm}$    0.013\\
12&   0.173 ${\color{darkred}\pm}$  0.027 &0.328 ${\color{darkred}\pm}$    0.047 \\
13&   0.151 ${\color{darkred}\pm}$  0.009 &0.272 ${\color{darkred}\pm}$    0.022\\
14&   0.157 ${\color{darkred}\pm}$  0.015 &0.322 ${\color{darkred}\pm}$    0.034\\
15&   0.147 ${\color{darkred}\pm}$  0.015 &0.261 ${\color{darkred}\pm}$    0.038\\
16&   0.161 ${\color{darkred}\pm}$  0.020 &0.332 ${\color{darkred}\pm}$    0.048\\
17&   0.183 ${\color{darkred}\pm}$  0.019 &0.300 ${\color{darkred}\pm}$    0.037	\\
18&   0.122 ${\color{darkred}\pm}$  0.037 &0.282 ${\color{darkred}\pm}$    0.057	\\
19&   0.132 ${\color{darkred}\pm}$  0.038 &0.292 ${\color{darkred}\pm}$    0.056	\\
20&   0.137 ${\color{darkred}\pm}$  0.028 &0.202 ${\color{darkred}\pm}$    0.095\\
21&   0.144 ${\color{darkred}\pm}$  0.039 &0.217 ${\color{darkred}\pm}$    0.075	\\
22&   0.153 ${\color{darkred}\pm}$  0.023 &0.300 ${\color{darkred}\pm}$    0.054\\
23&   0.152 ${\color{darkred}\pm}$  0.028 &0.302 ${\color{darkred}\pm}$    0.046	\\
24&   0.195 ${\color{darkred}\pm}$  0.027 &0.328 ${\color{darkred}\pm}$    0.049	\\
25&   0.184 ${\color{darkred}\pm}$  0.016 &0.303 ${\color{darkred}\pm}$    0.039\\
26&   0.177 ${\color{darkred}\pm}$  0.018 &0.416 ${\color{darkred}\pm}$    0.049\\
27&   0.149 ${\color{darkred}\pm}$  0.025 &0.326 ${\color{darkred}\pm}$    0.039		\\
28&   0.179 ${\color{darkred}\pm}$  0.024 &0.310 ${\color{darkred}\pm}$    0.039\\
29&   0.153 ${\color{darkred}\pm}$  0.016 &0.336 ${\color{darkred}\pm}$    0.036\\
30&   0.145 ${\color{darkred}\pm}$  0.026 &0.378 ${\color{darkred}\pm}$    0.054\\
31&   0.129 ${\color{darkred}\pm}$  0.014 &0.349 ${\color{darkred}\pm}$    0.032\\
32&   0.225 ${\color{darkred}\pm}$  0.024 &0.366 ${\color{darkred}\pm}$    0.041\\
33&   0.167 ${\color{darkred}\pm}$  0.013 &0.319 ${\color{darkred}\pm}$    0.029\\
34&   0.174 ${\color{darkred}\pm}$  0.027 &0.328 ${\color{darkred}\pm}$    0.047\\
  \hline
 \end{tabular}}
\end{table*}


\section{Spectral Analysis}
\label{sec4}
We  performed spectral  analysis  in the  0.5$-$10.0~keV energy  range using Xspec  12.8.2q and 12.10.0c \citep{1996Arnaud}.  X-ray spectra of \src have generally been interpreted with multiple blackbody functions.  Alford (2016) employed three blackbody functions for those observations performed during the first 2.5 years after the outburst and used two blackbodies for the rest. We tested the statistical necessity of adding a third blackbody to the data via Monte Carlo simulations and applying F-test. In particular, we generated 1000 simulated spectra using two blackbody functions as the seed model, and fit them with three blackbody functions, as well as two blackbody functions. We find that the change in  $\chi^2$ of the three blackbodies were not significant. Thus, we conclude that the addition of the third blackbody function is not essential. Therefore, we model the continuum X-ray spectra of \src with two blackbody functions: the cooler component is most likely originating from the whole neutron star surface,  while the hotter component is expected to originate from a small hot spot near the magnetic pole(s).

An accurate determination of interstellar hydrogen column density ({\it nH})
is  essential  in soft  X-ray  spectral  modeling, especially  when  a
blackbody   function   is   employed.    In  the   recent   study   of
\citet{Alford2016},   nH   was  obtained   as   $0.95\times10^{22}{\rm
  cm}^{-2}$ and kept  constant for all their  spectral analysis (using
      {\it wabs} in  Xspec). Here, we employed the  {\it tbabs} model,
      assuming the  ISM abundance  presented by  \citet{Wilms2000} and
      found  that  the  nH  values  vary  in  a  wide  range  covering
      $0.9-1.5\times10^{22}{\rm  cm}^{-2}$ from  2003 to  present.  In
      order to  obtain a  better constrain on  the this  parameter and
      other  spectral parameters,  we  fit the  initial 17  XMM-Newton
      observations simultaneously  by linking  their nH  parameter and
      obtain  a best  fit value  as $0.92{\pm  0.02 }\times10^{22}{\rm
        cm}^{-2}$. Assuming  that the interstellar  extinction remains
      stable, we  kept the  nH constant at  this value  throughout our
      spectral modeling.

      \begin{table*}[t]
        \begin{centering}
 \caption{Fit Results of Spectral Parameters}
 \label{tab:kt}
\setlength{\tabcolsep}{5.4pt}
\scriptsize{
 \begin{tabular}{cccccccc}
  \hline \Tstrut
  \# & kT$_1$ & Flux$_1^{\dagger}$ & kT$_2$  & Flux$_2^{\dagger}$ & Line & $\chi^2$ & dof\\
& (keV) &   &   (keV) &   \\
\hline\\
1&	 0.26 ${\color{darkred}\pm}$ 0.01 &		13.9 ${\color{darkred}\pm}$ 0.4 &    0.69 ${\color{darkred}\pm}$ 0.01 &    45.2 ${\color{darkred}\pm}$ 0.6 &   asym  & 118.8 &  110  \\
2&	 0.25 ${\color{darkred}\pm}$ 0.03 &		13.9 ${\color{darkred}\pm}$ 1.1 &    0.67 ${\color{darkred}\pm}$ 0.02 &    44.8 ${\color{darkred}\pm}$ 1.6 &   --    & 58.1  &  62	 \\
3&	 0.26 ${\color{darkred}\pm}$ 0.01 &		12.0 ${\color{darkred}\pm}$ 0.9 &    0.70 ${\color{darkred}\pm}$ 0.01 &    22.9 ${\color{darkred}\pm}$ 0.7 &   asym  & 120.1 &  107  \\
4&	 0.29 ${\color{darkred}\pm}$ 0.01 &	 	 8.6 ${\color{darkred}\pm}$ 0.3 &    0.70 ${\color{darkred}\pm}$ 0.01 &    11.3 ${\color{darkred}\pm}$ 0.2 &   asym  & 161.8 &  109   \\
5&	 0.23 ${\color{darkred}\pm}$ <0.01 &	 7.5 ${\color{darkred}\pm}$ 0.2 &    0.61 ${\color{darkred}\pm}$ 0.01 &     5.3 ${\color{darkred}\pm}$ 0.1 &    asym &  125.1 & 95	  \\
6&	 0.21 ${\color{darkred}\pm}$ <0.01 &	 6.0 ${\color{darkred}\pm}$ 0.1 &    0.56 ${\color{darkred}\pm}$ 0.02 &     1.6 ${\color{darkred}\pm}$ 0.1 &    asym &  66.6  & 66	  \\
7&	 0.19 ${\color{darkred}\pm}$ <0.01 &	 5.1 ${\color{darkred}\pm}$ 0.1 &    0.53 ${\color{darkred}\pm}$ 0.02 &     0.9 ${\color{darkred}\pm}$ <0.1 &  asym  &  56.2  & 54	  \\
8&	 0.18 ${\color{darkred}\pm}$ <0.01 &	 4.2 ${\color{darkred}\pm}$ 0.1 &    0.45 ${\color{darkred}\pm}$ 0.02 &     0.8 ${\color{darkred}\pm}$ 0.1 &    --   &  83.3  & 81	  \\
9&	 0.18 ${\color{darkred}\pm}$ <0.01 &	 4.4 ${\color{darkred}\pm}$ 0.1 &    0.38 ${\color{darkred}\pm}$ 0.02 &     0.7 ${\color{darkred}\pm}$ 0.1 &    sym  &    65.1   & 52	  \\
10& 0.18  ${\color{darkred}\pm}$ 0.01 &	     4.4 ${\color{darkred}\pm}$ 0.1 &    0.38 ${\color{darkred}\pm}$ 0.03 &     0.5 ${\color{darkred}\pm}$ 0.1 &   asym  & 37.2  & 45	  \\
11& 0.18  ${\color{darkred}\pm}$ 0.01 &	     4.4 ${\color{darkred}\pm}$ 0.1 &    0.36 ${\color{darkred}\pm}$ 0.02 &     0.6 ${\color{darkred}\pm}$ <0.1 & asym   & 87.6  & 60	  \\
12& 0.18  ${\color{darkred}\pm}$ 0.01 &	     4.0 ${\color{darkred}\pm}$ 0.1 &    0.41 ${\color{darkred}\pm}$ 0.04 &     0.5 ${\color{darkred}\pm}$ 0.1 &   asym  & 59.1  & 67	 \\
13& 0.19  ${\color{darkred}\pm}$ <0.01 &	 4.5 ${\color{darkred}\pm}$ 0.1 &    0.42 ${\color{darkred}\pm}$ 0.05 &     0.2 ${\color{darkred}\pm}$ 0.1 &    asym & 80.4  & 51	 \\
14& 0.19  ${\color{darkred}\pm}$ 0.01 &	     4.4 ${\color{darkred}\pm}$ 0.1 &    0.50 ${\color{darkred}\pm}$ 0.08 &     0.2 ${\color{darkred}\pm}$ 0.1 &    asym & 65.1  & 34	 \\
15& 0.19  ${\color{darkred}\pm}$ 0.01 &	     4.2 ${\color{darkred}\pm}$ 0.2 &    0.40 ${\color{darkred}\pm}$ 0.05 &     0.3 ${\color{darkred}\pm}$ 0.1 &    asym & 66.2  & 34	 \\
16& 0.17  ${\color{darkred}\pm}$ <0.01 &	 4.1 ${\color{darkred}\pm}$ 0.1 &    0.37 ${\color{darkred}\pm}$ 0.04 &     0.4 ${\color{darkred}\pm}$ 0.1 &    --   & 47.0  & 34	 \\
17& 0.18  ${\color{darkred}\pm}$ 0.01 &	     4.2 ${\color{darkred}\pm}$ 0.1 &    0.41 ${\color{darkred}\pm}$ 0.04 &     0.5 ${\color{darkred}\pm}$ 0.1 &    sym  & 66.9  & 60	 \\
18& 0.18  ${\color{darkred}\pm}$ 0.01 &	     4.0 ${\color{darkred}\pm}$ 0.2 &    0.49 ${\color{darkred}\pm}$ 0.55 &     0.5 ${\color{darkred}\pm}$ 0.1 &    --   & 28.1  & 29	 \\
19& 0.17  ${\color{darkred}\pm}$ 0.01 &	     4.0 ${\color{darkred}\pm}$ 0.2 &    0.42 ${\color{darkred}\pm}$ 0.15 &     0.6 ${\color{darkred}\pm}$ 0.1 &    --   & 26.9  & 30	 \\
20& 0.18  ${\color{darkred}\pm}$ 0.01 &	     4.8 ${\color{darkred}\pm}$ 0.3 &    0.42 ${\color{darkred}\pm}$ 0.12 &     0.3 ${\color{darkred}\pm}$ 0.2 &    asym & 29.0  & 26	\\
21& 0.17  ${\color{darkred}\pm}$ 0.01 &	     4.3 ${\color{darkred}\pm}$ 0.2 &    0.47 ${\color{darkred}\pm}$ 0.13 &     0.5 ${\color{darkred}\pm}$ 0.1 &    --   & 26.9  & 27	\\
22& 0.18  ${\color{darkred}\pm}$ 0.01 &	     4.5 ${\color{darkred}\pm}$ 0.1 &    0.51 ${\color{darkred}\pm}$ 0.44 &     0.2 ${\color{darkred}\pm}$ 0.1 &    sym  & 29.0  & 26	\\
23& 0.17  ${\color{darkred}\pm}$ 0.01 &	     4.2 ${\color{darkred}\pm}$ 0.2 &    0.46 ${\color{darkred}\pm}$ 0.03 &     0.5 ${\color{darkred}\pm}$ 0.1 &    asym & 30.2  & 33	\\
24& 0.17  ${\color{darkred}\pm}$ <0.01 &	 4.3 ${\color{darkred}\pm}$ 0.2 &    0.52 ${\color{darkred}\pm}$ 0.19 &     0.4 ${\color{darkred}\pm}$ 0.1 &    --   & 47.3  & 34	\\
25& 0.19  ${\color{darkred}\pm}$ 0.01 &	     4.4 ${\color{darkred}\pm}$ 0.1 &    0.42 ${\color{darkred}\pm}$ 0.10 &     0.2 ${\color{darkred}\pm}$ 0.1 &    asym  & 40.2  & 33	 \\
26& 0.17  ${\color{darkred}\pm}$ 0.01 &	     4.8 ${\color{darkred}\pm}$ 0.1 &    0.42 ${\color{darkred}\pm}$ 0.08 &     0.3 ${\color{darkred}\pm}$ 0.1 &   asym  & 26.9  & 30	\\
27& 0.17  ${\color{darkred}\pm}$ 0.01 &	     4.0 ${\color{darkred}\pm}$ 0.1 &    0.38 ${\color{darkred}\pm}$ 0.05 &     0.5 ${\color{darkred}\pm}$ 0.1 &     --  & 45.0  & 48	 \\
28& 0.18  ${\color{darkred}\pm}$ 0.01 &	     4.3 ${\color{darkred}\pm}$ 0.1 &    0.43 ${\color{darkred}\pm}$ 0.06 &     0.4 ${\color{darkred}\pm}$ 0.1 &     --  & 145.4 & 93	\\
29& 0.17  ${\color{darkred}\pm}$ <0.01 &	 4.4 ${\color{darkred}\pm}$ 0.1 &    0.38 ${\color{darkred}\pm}$ 0.05 &     0.4 ${\color{darkred}\pm}$ 0.1 &    --   & 105.7 & 96	\\
30& 0.17  ${\color{darkred}\pm}$ 0.01 &	     4.2 ${\color{darkred}\pm}$ 0.1 &    0.35 ${\color{darkred}\pm}$ 0.12 &     0.4 ${\color{darkred}\pm}$ 0.1 &   --    & 70.5  & 54	 \\
31& 0.17  ${\color{darkred}\pm}$ <0.01 &	 4.4 ${\color{darkred}\pm}$ 0.1 &    0.37 ${\color{darkred}\pm}$ 0.03 &     0.4 ${\color{darkred}\pm}$ 0.1 &  --     & 145.7 & 123 \\
32& 0.18  ${\color{darkred}\pm}$ <0.01 &	 4.1 ${\color{darkred}\pm}$ 0.2 &    0.36 ${\color{darkred}\pm}$ 0.01 &     0.6 ${\color{darkred}\pm}$ 0.2 &    --   & 59.1  & 86	 \\
33& 0.17  ${\color{darkred}\pm}$ <0.01 &	 4.1 ${\color{darkred}\pm}$ 0.1 &    0.31 ${\color{darkred}\pm}$ 0.02 &     0.6 ${\color{darkred}\pm}$ 0.1 &  --     & 161.5 & 126 \\
34& 0.17  ${\color{darkred}\pm}$ 0.01 &	     4.6 ${\color{darkred}\pm}$ 0.2 &    0.39 ${\color{darkred}\pm}$ 0.04 &     0.5 ${\color{darkred}\pm}$ 0.1 &   --	   & 90.3  & 89	\\
  \hline 
 \end{tabular}}
 \footnotesize{
   \begin{flushleft}
     $^{\dagger}$ Unabsorbed 0.5$-$10.0~keV flux in units of 10$^{-12}$ erg cm$^{-2}$ s$^{-1}$.  
\end{flushleft} }
 \end{centering}
\end{table*}

\begin{figure*}[t]
  \begin{centering}
 \subfloat{\includegraphics[scale=0.3]{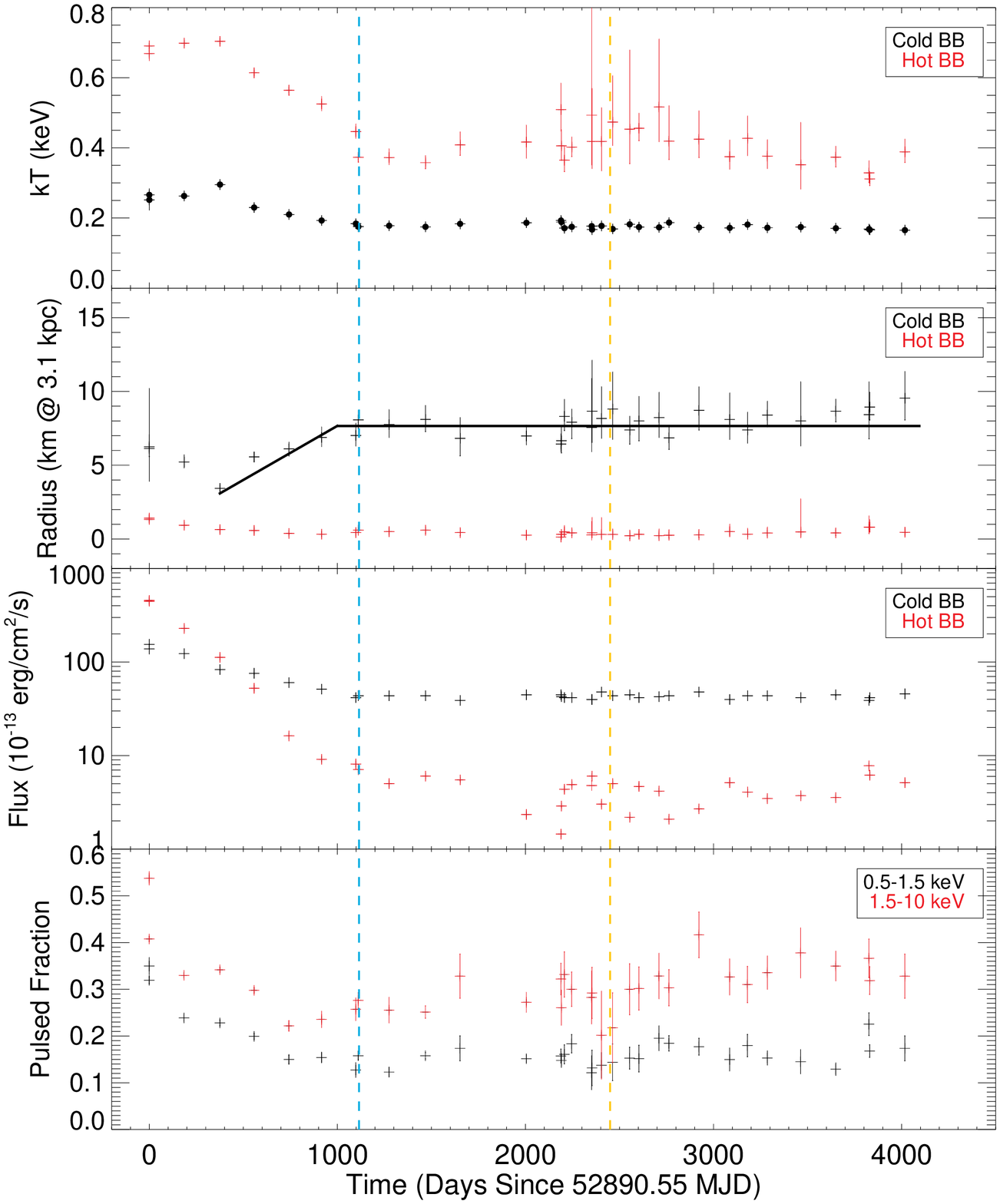}}
 \qquad\subfloat{}
  \subfloat{\includegraphics[scale=0.4]{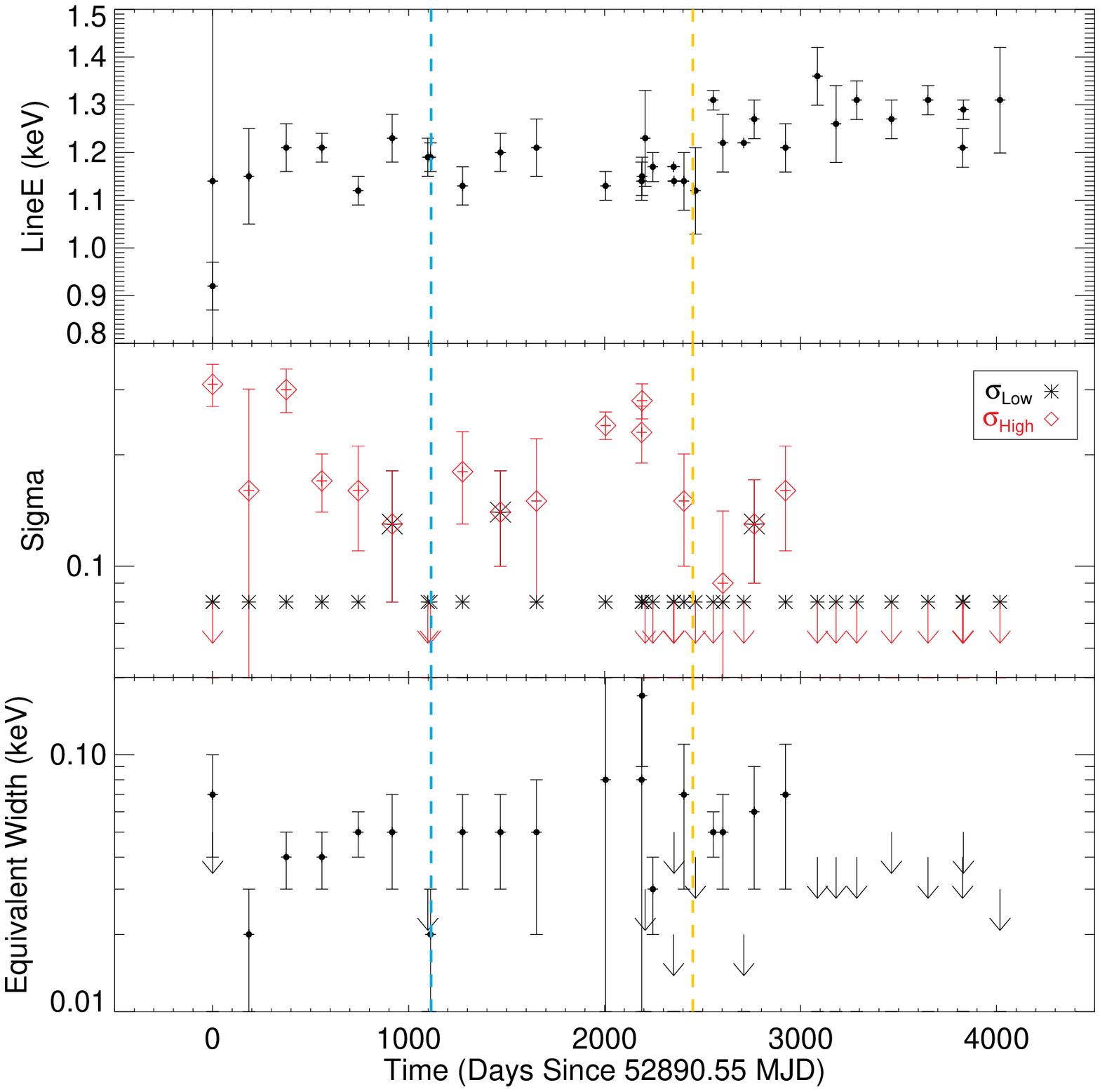}}
 \caption{ {\it \bf Left :} Long-term  evolution of  the spectral model  parameters and
pulsed fractions.  {\it Top panel:}  Evolution of the cold  and hot
   blackbody temperatures, {\it Second  panel from top:} Variations of
   the inferred blackbody emitting radii  for the two components, {\it
     Third panel  from top:} Evolution  of the flux of  each blackbody
   component,  and  {\it  Bottom   panel:}  Evolution  of  the  pulsed
   fractions in the 0.5$-$1.5~keV  energy range (black) and 1.5-10~keV
   (red) range. {\it \bf Right :} Long-term  evolution of  the absorption line parameters. {\it First  panel:}  Long-term evolution  of  the  centroid
   energy  of absorption  line energy.  {\it Second  panel:} Left  and
   right widths of its asymmetric  shape. {\it Last panel:} Equivalent
   width of the absorption line. The upside down arrow indicates upper
   limits.  The  vertical  dashed  blue  line  indicate  the  time  of
   frequency change,  and the  dashed yellow  line corresponds  to the
   time of possible anti-glitch \citep{Pintore2016}.}
 \label{fig:result}
 \end{centering}
\end{figure*}

We found that all 34 Chandra and XMM-{\it Newton} spectra of \src were
well fit with the assumed model  of two blackbodies plus an asymmetric
gaussian absorption:  these fits yield reduced-$\chi^2$  values mainly
between  0.85   and  1.4   (see \autoref{tab:kt}   for  individual
values).   Temperatures  of   the  two   blackbody  functions   evolve
significantly during the  first 1000 days of the  outburst. The cooler
blackbody  component  (kT$_1$) started  with  0.26  keV and  showed  a
marginal increase to its maximum value  of 0.29 keV then declined down
to about 0.19 keV in 200 days, and remained around this constant level
since  then. A  very similar  trend is  seen in  the hotter  blackbody
component (kT$_2$),  which was  at 0.69  keV at  the onset,  showing a
marginal  increase to  the  maximum value  of 0.70  keV  in about  400
days. kT$_2$  then declined monotonically  down to 0.5 keV  in another
700 days. Since then it  remained approximately constant at this value
(see the  first panel of \autoref{fig:result}, and \autoref{tab:kt}
for details).

We also calculated  the radii of blackbody emitting  regions by taking
the distance to  the \src as $3.1\pm 0.5$  kpc \citep{Durant2006} into
account. The  radius of the  cold blackbody component (R$_1$),  on the
other hand, exhibits  an intriguing trend: it declined from  6 to 3~km
in about  400 days after the  outburst onset. Then, R$_1$  expanded to
about  7~km in  another  400  days (see  the  second  panel of
\autoref{fig:result}).  The emitting  radius of  the cold  component remained
constant since then. The region  of the hot component (R$_2$) declined
marginally from 0.26 to 0.20 km at the very beginning of the outburst,
then remained constant at this level for almost a decade. It is likely
that the location of this hot spot is near the magnetic poles.

We modeled  the time evolution of  the radius of the  cooler blackbody
component starting from  about 300 days following  the outburst onset,
using two  linear trends (see  the solid line  in the second  panel of \autoref{fig:result}).  We found  that the  rise of  the radius  of the
cooler blackbody component stops at around 1000 days with a maximum of
about  7.5~km. The  slope of  the linear  trend indicates  a speed  of
spreading. After the break, the radius of this component is consistent
with being nearly constant till the latest observation.

We present unabsorbed  X-ray flux values in  the 0.5$-$10.0~keV range,
in the third panel of \autoref{fig:result}. Flux of the colder blackbody component
(F$_1$) decreased  slightly at the  early phases of the  outburst, and
remains     constant      at     around     $4\times10^{-12}~\rm{erg}~
\rm{cm}^{-2}~\rm{s}^{−1}$. Flux of the hot blackbody component (F$_2$)
shows    an    exponential    decay    from    $5\times10^{-11}$    to
$10^{-12}~\rm{erg}~\rm{cm}^{-2}~\rm{s}^{−1}$  during  its  first  1000
days. The cooling  process is likely responsible  for this significant
decrease. In the  rest of the observations,  F$_2$ fluctuates around
$10^{-12}~\rm{erg}~\rm{cm}^{-2}~\rm{s}^{−1}$. Therefore,  the ratio of
F$_1$ to F$_2$ has been steady at around 4 for almost a decade.


 \section{Absorption Line}
 \label{sec5}
 The absorption  line around 1.1~keV has  been reported in a  number of
studies  of   \src  \citep{Alford2016,2011bernardini,Camilo2016},  and
vaguely  attributed to  resonant cyclotron  processes in  the strongly
magnetized magnetosphere.  We also detected  this feature in 19 out of
34 observations. It is important to note that the observed absorption feature is  not always symmetric. When detected, we found that this feature is  often asymmetric; it
can be  represented with a  symmetric Gaussian  profile only in  a few
cases.   To  accurately  represent  this  feature,  we  introduced  an
asymmetric  Gaussian profile  where the  widths of  the feature  is at
lower  and higher  energies of  its centroid  are not  necessarily the
same.  

\begin{figure*}[t]
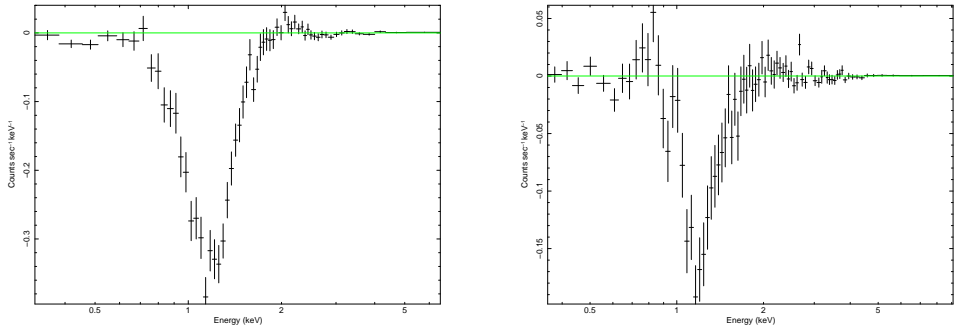

  \begin{centering}
 \subfloat{\includegraphics[scale=0.25,angle=270]{0301270301.eps}}
 \qquad\subfloat{}
  \subfloat{\includegraphics[scale=0.25,angle=270]{0301270401.eps}}
 \caption{Sample residuals showing symmetric and asymmetric absorption line feature of XMM-Newton observations with ID: 0301270301 {\it (left)} and 0301270401 {\it (right)}.}
 \label{fig:difline}
\end{centering}
\end{figure*}

In \autoref{fig:xmm}, we  present  an  example  case  where  the  asymmetric  shape  of  the
absorption feature can  be seen clearly. To account for  this fact, we
include an asymmetric Gaussian  function whose lower-energy half-width
can  be different  than  its higher-energy  half-width.  To model  the
detected absorption  line, we  construct an asymmetric  Gaussian
model  function  in  {\it  Xspec}  that resembles  the  shape  of  the
absorption feature. The  absorption feature is observed  to exhibit in
certain cases a sharp decay followed by a relatively slow recovery. We
define the model as
\begin{eqnarray}
\label{eq:3}
\begin{aligned}
A(E) = \begin{cases}
K e^{\left(\frac{-(E-E_c)^2}{2 \sigma_1^2}\right)}, &       E < E_c     \\
K e^{\left(\frac{-(E-E_c)^2}{2 \sigma_2^2}\right)}, &        E \geq E_c
\end{cases}
\end{aligned}
\end{eqnarray}
\\
where $K$  is the normalization, $E_c$  is the centroid energy  of the feature, $\sigma_1$ and $\sigma_2$ are the left and right width of the Gaussian shape,  respectively. In  each spectrum where  the absorption feature is present,  we found that the $\sigma_1$ is  always less than $\sigma_2$. We present the cases of a symmetric and an asymmetric feature in \autoref{fig:difline}. In the remaining 15 observations,  we could only obtain  upper limits to the either  or both  half-widths of  the line feature  and had  to set their value to the instrumental spectral resolution limit. When fitting the line, it is not always possible to calculate the lower-energy half-width of the absorption line S$_1$ due to the limits of the energy range. Thus, we keep S$_1$ is fixed at 0.08 due to the minimum of the energy resolution of Chandra.

Another important issue to note is the shift in the centroid energy of the line. Apart  from  the  very  first  few  observation, the centroid energy of the line resides in a narrow range of $1.1-1.3$~keV (see the  top panel of \autoref{fig:result} and \autoref{tab:line}  for details).   

The last two columns of \autoref{tab:line} list $\sigma_1^s$ and $\sigma_2^s$ of the best fit absorption feature obtained simultaneously. We found that $\sigma_2^s$ parameter in observations 1, 4, 13, 14, and 15 are consistent one another, and significantly larger than $\sigma_1^s$. To better constrain $\sigma_2^s$ for these observations, we fitted them simultaneously by fixing the $\sigma_1^s$ to 0.08 and linking the $\sigma_2^s$ parameter. In the joint analysis, we allowed $kT_1$ and $kT_2$ for each observation float independently. As a result of this joint investigation, we determine $\sigma_2^s$ for this group of five observations as 0.27 $\pm$ 0.02 keV. Similarly, $\sigma_2^s$ of observations of 5, 6, 7, 10, and 11 are relatively larger than $\sigma_1^s$. Joint spectral investigations of these observations yield an $\sigma_2^s$ of 0.17 $\pm$ 0.02 keV for this group. Finally, $\sigma_2^s$ in observations 7, 11, and 25 are marginally larger than $\sigma_1^s$. The joint spectral fit results in 0.13 $\pm$ 0.02 keV for them.

The  equivalent  width   of  the absorption line feature is also shown \autoref{fig:result} and indicates an overall variability which is further discussed in the next section.  Note that in  a few cases  the equivalent width  of the feature could not be constrained.

\begin{table*}[t]
  \centering
 \caption{Fit Results of Line Parameters}
 \label{tab:line}
\setlength{\tabcolsep}{5.4pt}
\scriptsize{
 \begin{tabular}{ccccccc}
  \hline \Tstrut
  \# & Line Energy & $\sigma_1$ & $\sigma_2$  & Equivalent Width  & $\sigma_1^s$   & $\sigma_2^s$  \\
  & (keV) & (keV)  & (keV)  & (keV) & (keV)  & (keV) \\
\hline\\
1&	  0.92 ${\color{darkred}\pm}$ 0.05 & 	0.08  	    	&   0.31 ${\color{darkred}\pm}$ 0.04   	    &  0.07	${\color{darkred}\pm}$ 0.03	 & 0.08   & 0.27  ${\color{darkred}\pm}$ 0.02  \\
2&	  1.14 ${\color{darkred}\pm}$ 0.85 &	0.08			&   0.08                         			&  $<$0.05                                  \\
3&	  1.15 ${\color{darkred}\pm}$ 0.10 & 	0.08	   		&   0.16 ${\color{darkred}\pm}$ 0.14    	&  0.02 ${\color{darkred}\pm}$ 0.01        \\
4&	  1.21 ${\color{darkred}\pm}$ 0.05 & 	0.08	    	&	0.30 ${\color{darkred}\pm}$ 0.04    	&  0.04 ${\color{darkred}\pm}$ 0.01  & 0.08   & 0.27  ${\color{darkred}\pm}$ 0.02  \\
5&	  1.21 ${\color{darkred}\pm}$ 0.03 & 	0.08	    	&	0.17 ${\color{darkred}\pm}$ 0.03    	&  0.04 ${\color{darkred}\pm}$ 0.01  & 0.08   & 0.17  ${\color{darkred}\pm}$ 0.02  \\
6&	  1.12 ${\color{darkred}\pm}$ 0.03 & 	0.08      		&	0.16 ${\color{darkred}\pm}$ 0.05    	&  0.05 ${\color{darkred}\pm}$ 0.01  & 0.08   & 0.17  ${\color{darkred}\pm}$ 0.02  \\
7&	  1.23 ${\color{darkred}\pm}$ 0.05 & 	0.13 ${\color{darkred}\pm}$ 0.05 & 0.13 ${\color{darkred}\pm}$ 0.05 &  0.05 ${\color{darkred}\pm}$ 0.02 & 0.08   & 0.17, 0.13  ${\color{darkred}\pm}$ 0.02   	\\
8&	  1.19 ${\color{darkred}\pm}$ 0.04 &    0.08     		&	0.08                             		&  $<$0.03 \\
9&	  1.19 ${\color{darkred}\pm}$ 0.03 & 	0.08  			&   0.08                         			&  0.02 ${\color{darkred}\pm}$ 0.01    \\
10&   1.13 ${\color{darkred}\pm}$ 0.04 & 	0.08 			&   0.18 ${\color{darkred}\pm}$ 0.05    	&  0.05 ${\color{darkred}\pm}$ 0.02  & 0.08   & 0.17  ${\color{darkred}\pm}$ 0.02  \\
11&   1.20 ${\color{darkred}\pm}$ 0.04 & 	0.14 ${\color{darkred}\pm}$ 0.04 &  0.14 ${\color{darkred}\pm}$ 0.04   &  0.05 ${\color{darkred}\pm}$ 0.02  & 0.08   & 0.17, 0.13  ${\color{darkred}\pm}$ 0.02  \\
12&   1.21 ${\color{darkred}\pm}$ 0.06 & 	0.08 	    	&   0.15 ${\color{darkred}\pm}$ 0.07  	    &  0.05 ${\color{darkred}\pm}$ 0.03     \\
13&   1.13 ${\color{darkred}\pm}$ 0.03 & 	0.08 	    	&   0.24 ${\color{darkred}\pm}$ 0.02 		&  0.08 ${\color{darkred}\pm}$ 0.12 & 0.08   & 0.27  ${\color{darkred}\pm}$ 0.02 \\
14&   1.14 ${\color{darkred}\pm}$ 0.04 & 	0.08    		&   0.23 ${\color{darkred}\pm}$ 0.04  	    &  0.08 ${\color{darkred}\pm}$ 0.14 & 0.08   & 0.27  ${\color{darkred}\pm}$ 0.02  \\
15&   1.15 ${\color{darkred}\pm}$ 0.04 & 	0.08    		&   0.28 ${\color{darkred}\pm}$ 0.03  	    &  0.17 ${\color{darkred}\pm}$ 0.08 & 0.08   & 0.27  ${\color{darkred}\pm}$ 0.02 \\
16&   1.23 ${\color{darkred}\pm}$ 0.10 &    0.08        	&   0.08                        	        &  $<$0.03			 \\
17&   1.17 ${\color{darkred}\pm}$ 0.03 & 	0.08 			&   0.08                     		    	&  0.03 ${\color{darkred}\pm}$ 0.01  \\
18&   1.17                      		&	0.08    		&	0.08                     	  		    &  $<$0.02 \\ 
19&   1.14                      	   &    0.08    		&	0.08                      		        &  $<$0.05 	\\	
20&   1.14 ${\color{darkred}\pm}$ 0.06 & 	0.08 			&   0.15 ${\color{darkred}\pm}$ 0.05 	 	&  0.07 ${\color{darkred}\pm}$ 0.04    \\
21&   1.12 ${\color{darkred}\pm}$ 0.09 &	0.08    		&	0.08                         	   		&   $<$0.04		\\
22&   1.31 ${\color{darkred}\pm}$ 0.02 & 	0.08 			&   0.08                         			&  0.05 ${\color{darkred}\pm}$ 0.01   		\\
23&   1.22 ${\color{darkred}\pm}$ 0.06 & 	0.08 			&   0.09 ${\color{darkred}\pm}$ 0.05 	 	&  0.05 ${\color{darkred}\pm}$ 0.02    \\
24&   1.22                           &  	0.08   		    &   0.08                         	 		&  $<0.02$	\\
25&   1.27 ${\color{darkred}\pm}$ 0.04 & 	0.13 ${\color{darkred}\pm}$ 0.04  &   0.13 ${\color{darkred}\pm}$ 0.04 	 	&  0.06 ${\color{darkred}\pm}$ 0.03 & 0.08   & 0.13  ${\color{darkred}\pm}$ 0.02  \\
26&   1.21 ${\color{darkred}\pm}$ 0.05 & 	0.08        	&   0.16 ${\color{darkred}\pm}$ 0.05 	 	&  0.07 ${\color{darkred}\pm}$ 0.04  \\
27&   1.36 ${\color{darkred}\pm}$ 0.06	&	0.08      		&	0.08                        			&  $<$0.04 	    \\  
28&   1.26 ${\color{darkred}\pm}$ 0.08	&	0.08      		&	0.08                        			&  $<$0.04	   	\\
29&   1.31 ${\color{darkred}\pm}$ 0.04 &	0.08      		&	0.08                        			&  $<$0.04	   	\\
30&   1.27 ${\color{darkred}\pm}$ 0.04	&	0.08      		&	0.08                        			&  $<$0.05	   	\\  
31&   1.31 ${\color{darkred}\pm}$ 0.03	&   0.08   			&	0.08                        			&  $<$0.04	   	 \\  
32&   1.21 ${\color{darkred}\pm}$ 0.04 &	0.08      		&	0.08                        			&  $<$0.04	   	 \\
33&   1.29 ${\color{darkred}\pm}$ 0.02	&	0.08    		&	0.08                        			&  $<$0.05 	   	 \\	 
34&   1.31 ${\color{darkred}\pm}$ 0.11	&	0.08      		&	0.08                        		 	&  $<$0.03	   	\\
  \hline
 \end{tabular}}
 \footnotesize{
   \begin{flushleft}
   $\sigma_1^s$ and $\sigma_2^s$ represents the values obtained from simultaneous fitting.\\ 
\end{flushleft} }
\end{table*}\

Overall, we find that, when detected, spectral properties of the absorption feature are highly variable, and it cannot be detected in 15 observations. To better understand whether the absence of the absorption line in these 15 observations is due to the insufficient duration or decline in its X-ray flux, we have performed simulations with Xspec{\footnote{https://heasarc.gsfc.nasa.gov/xanadu/xspec/manual/node73.html}}. Using the continuum model employed (2 blackbody functions and an asymmetric Gaussian absorption line), we varied the X-ray flux for the values of these 15 observations, ranging from $8.0\times10^{-13}$ to $3.0\times10^{-11}$ erg cm$^{-2}$ s$^{-1}$, and set exposure times at 20 ks, 40 ks and 50 ks. We found that the absorption feature is always detectable if the exposure of the observation is as large as 50 ks. Therefore, we conclude that the absence of this absorption line is not related to any instrumental effects.


 \section{Discussion and Conclusions}
\label{sec6}
We  investigated long-term  spectral  and temporal  evolution of  \src
using X-ray  data collected  over a  long time  span with  Chandra and
XMM-{\it Newton}. We  modeled surface thermal evolution  of \src using
two  blackbody components  of which  one is  most likely  arising from
nearly the entire neutron star surface at a temperature of 0.2 keV and
the other component  is a hot spot  at 0.6~keV on on  the neutron star
surface. We found that the radius of the emitting region of the cooler
component is  about 7~km, while  that of the  hot spot on  the neutron
star is  about 0.2~km (assuming that  the distance of \src  is $3.1\pm
0.5$~kpc  \citep{Durant2006}). Once a magnetar outburst is
ignited, the release of internal  magnetic energy heats up the neutron
star crust most likely near  or at the magnetic pole \citep{2009Belo}.
The excess heat  energy on small portion of the  surface (hot spot) is
expected  to  spread  and  heat  up  a  much  larger  portion  of  the
surface. We found that the energy of the heated spot on the surface of
\src dissipates on  a timescale of about 3 years,  after which neither
the temperatures,  nor the  sizes of emitting  radii of  both emission
components  change  significantly.  The  X-ray spectra  of  \src  were
previously  modeled  with slightly  different  models,  such as two blackbody 
\citep{Gotthelf2007} or three blackbody components \citep{Alford2016,Bernardini2009}.   
Based on our detailed numerical  investigations, we found that  statistically there
is no requirement for a third blackbody component.

We identified  two distinct emission  regions; a  hot spot and  a much
wider surface,  which cools down on  a much slower pace.  The hot spot
occupies about 0.8\% of the neutron  star surface area, and cools down
faster while  its size  remain fairly constant.  This is  in agreement
with the fact that the imparted heat  energy to a small section on the
surface at the  onset of the outburst spreads to  the whole surface of
the neutron star in a time scale of a few years.

We also investigated any possible correlations between spectral and temporal properties in  order to better understand the link between these quite related domains
using two methods: A cross-correlation  scheme and  $Z-Transform$ $Discrete$  $Correlation$ $Function$  $(ZDCF)$ \citep{Alexander2013}.  The latter method yields the parameter errors as well. For the pulsed fractions, we employed those in the 0.5-1.5 ($PF_{low}$) keV and 1.5-10.0 ($PF_{high}$) keV energy range. For this  examination, we considered  the data for   the   first  1500   days. We find a strong positive correlation between the temperature of the cool blackbody and pulsed fraction in the lower energy band while the correlation between the radius of the cool blackbody is lower than the hot blackbody (see \autoref{tab:cor}    for details). Temperature of the hot component shows a negative correlation with the pulsed fraction in the 1.5-10.0 keV band. We also computed the time lag between variations of spectral parameters and the pulsed fraction. Time lag between two parameters is important for testing models of continuum emission. Interestingly, there is a time lag between the temperatures of the blackbody components and their pulsed fraction order of days as seen in \autoref{tab:cor}. The fluxes of the blackbody components also correlate with their pulsed fraction.

\citet{2001zane} points out that an absorption line at the proton cyclotron energy may exist where the line equivalent width is between 30 eV and 1 keV and the line center is located at $0.5-5.0$ keV for field strengths 10$^{14}$-10$^{15}$ G. Cyclotron lines are entirely important for neutron stars because of not only give a direct measurement of the magnetic field but also give information about the physical conditions of the radiating regions \citep{1991Pavlov}. We rigorously studied the absorption line in various ways as shown in \S\ref{sec5}. We measured that the absorption line center is located around $\approx1.1$ keV. Note that the line centroid shifts in energy and its profile varies between being symmetric and asymmetric. According to \citet{2001Ozel, 2003Ozel}, the combination of vacuum polarization and cyclotron energy may cause such
asymmetric features. In the case of strong magnetic fields, vacuum polarization affects the polarization of normal modes, alter  the dielectric tensor, as well as the magnetic permeability tensor. Moreover, it also changes the interaction cross sections in the plasma.   

The  observed variability of the absorption feature can also be related to varying optical depth
in   the  magnetosphere   of XTE~J1810$-$197. \citet{2006Lyutikov},
\citet{Guver2007} and \citet{2007Fernandez} showed that the equivalent width of a  proton cyclotron line feature generated  in the atmosphere of  a  magnetar  can  be  affected  by  the  optical   depth  in  the magnetosphere.  We conclude that the variation of the equivalent width of the line could be attributed to variations in optical depth in the magnetosphere, likely due to loading or unloading particles in the line of sight.  Therefore, the line feature might get smeared in the  case of higher particle density,  while reappears when the density drops. \citet{2003Bignami} also stress that in the cyclotron scattering above the surface, cross-section depends on the angle between observer and magnetic axis. This yields to a variation of the feature depth. The line-width also vary if the observer's line of sight and the magnetic field are perpendicular. 

In a  recent study, \citet{Pintore2016} demonstrated  that there exist
timing irregularities  in the spin  period evolution of \src  the one
around MJD 55400 was interpreted as  an anti-glitch. They also noted a
frequency change around MJD 54000. On  the other hand, we marked these
particular  times in \autoref{fig:result}.  We find  that the  time of
frequency  change is  somehow  coincident with  spectral and  temporal
alterations for both hot and cold  BB components. We find no radiative
changes in it  emission at the time of suggested  anti-glitch. This is
consistent   with   the   anti-glitch  observed   from   1E   1841-045
\citep{2014sinem},  while  both  anti-glitch sources  differ  from  1E
2259+586  in which  the  anti-glitch was  associated  with clear  flux
enhancement \citep{2013Archibald}.

\begin{table*}[h]
  \centering
 \caption{Correlations between Pulsed Fractions and Spectral Parameters for the first 1500 days}
 \label{tab:cor}
\setlength{\tabcolsep}{5.4pt}
{\renewcommand{\arraystretch}{1.5}
\scriptsize{
 \begin{tabular}{ccccc}
  \hline \Tstrut
  Parameter Pairs & ZDCF & Time Lag (days) & Correlation & Time Lag (days) \\
 & (0.5-10.0 keV) &    (0.5-10.0 keV) & Coefficient &  \\
\hline
$kT_1$ -- $PF_{low}$  &  0.90$_{-0.04}^{+0.04}$    &  312$_{-71}^{+55}$ 	&	0.775  	&	-360	\\

$R_1$ -- $PF_{low}$	  &  -0.77$_{-0.08}^{+0.09}$   & 312$_{-55}^{+70}$ 	&	-0.859	&	-360	\\

$F_1$ -- $PF_{low}$	  &  0.95$_{-0.04}^{+0.03}$   &  26$_{-27}^{+24}$		&	0.941	&	0	\\

$kT_2$ -- $PF_{high}$  &  -0.70$_{-0.11}^{+0.13}$   & 693$_{-40}^{+67}$		&	0.695	&	-360	\\

$R_2$ -- $PF_{high}$  &  0.79$_{-0.14}^{+0.10}$   &  26$_{-27}^{+24}$		&	0.935	&	0	\\

$F_2$ -- $PF_{high}$  &  0.84$_{-0.11}^{+0.08}$   &  26$_{-27}^{+24}$		&	0.846 	&	0	\\ 
  \hline
 \end{tabular}}}
 
\end{table*}

\section{Acknowledgements}
We thank the anonymous referee for constructive comments. EV thanks Armin Vahdat Motlagh for his help and useful discussions. We thank Muhammed Diyaddin Ilhan for his help in the ZDCF analysis. This work was  supported by  Scientific  Research Project  Coordination Unit  of Istanbul University  with project numbers:  57462, 48934. EV  has been
supported  by  the  Scientific   and  Technological  Research  Council
(TUBITAK) through project 113F270. This  research has made use of data
obtained from  the High  Energy Astrophysics Science  Archive Research
Center (HEASARC), provided by NASA's Goddard Space Flight Center.

{}

\label{lastpage}

\end{document}